\documentclass[traditabstract]{aa}
\usepackage{graphicx}
\usepackage{aalongtable}
\usepackage{lscape}
\usepackage{txfonts}
\usepackage{natbib}
\bibpunct{(}{)}{;}{a}{}{,}
\usepackage{color}
\newcommand{\eg}{{\em e.g.}}

\begin{document}
\title{The red-sequence of 72 WINGS local galaxy clusters}

   \author{
 T. Valentinuzzi\inst{1},
 B.M. Poggianti \inst{2}, 
 G. Fasano\inst{2} \and
 M. D'Onofrio\inst{1} \and
 A. Moretti\inst{2} \and
 M. Ramella\inst{3} \and
 A. Biviano\inst{3} \and
 J. Fritz\inst{4} \and
 J. Varela\inst{2} \and
 A. Cava\inst{5} \and
 D. Bettoni\inst{2} \and
 B. Vulcani\inst{2,1} \and
 M. Moles\inst{6} \and
 W.J. Couch\inst{7}\and
 A. Dressler\inst{8}\and
 P. Kj{\ae}rgaard\inst{9}
 A. Omizzolo\inst{2}
          }

   \institute{
     Astronomy Department, University of Padova, Vicolo Osservatorio 2, 35122 Padova, Italy \and  
     INAF -- Padova Astronomical Observatory, Vicolo Osservatorio 5, 35122 Padova, Italy \and
     INAF -- Trieste Astronomical Observatory, via Tiepolo 11, 34131 Trieste, Italy \and
     Sterrenkundig Observatorium Vakgroep Fysica en Sterrenkunde Universeit Gent, Gent, Belgium \and   
     Departamento de Astrofisica, Universidad Computense de Madrid, Spain  \and      Centro de Estudios de Física del Cosmos de Aragón, Plaza de San Juan 1, 44001 Teruel, Spain \and 
     Centre for Astrophysics \& Supercomputing, Swinburne University, Hawthorn 3122, VIC, Australia\and
     Observatories of the Carnegie Institution of Washington, Pasadena, CA 91101, USA \and
     The Niels Bohr Institute, Juliane Maries Vej 30, 2100 Copenhagen, Denmark
    }

   \date{\today}


\abstract{
We study the color-magnitude red sequence and blue fraction of 72
X-ray selected galaxy clusters at $z=0.04-0.07$ from the WINGS survey,
searching for correlations between the characteristics of the red sequence
and the environment.
We consider the slope and scatter of the red sequence,
the number ratio of red luminous-to-faint galaxies, the blue fraction
and the fractions of ellipticals, S0s and spirals that compose the red 
sequence. None of these quantities correlate with the cluster 
velocity dispersion, X-ray luminosity, number of cluster substructures,
BCG prevalence over next brightest galaxies and spatial concentration
of ellipticals. Instead, the properties of the red sequence
depend strongly on local galaxy density. Higher density regions 
have a lower RS scatter, a higher luminous-to-faint ratio, a lower blue 
fraction, and a lower spiral fraction on the RS.
Our results highlight the prominent effect of the local density
in setting the epoch when galaxies
become passive and join the red sequence,
as opposed to the mass of the galaxy host structure.}


\authorrunning{Valentinuzzi et al.}

\titlerunning{Red-sequence of WINGS clusters}

\maketitle

%


\section{Introduction}

The location of galaxies in a color-absolute magnitude diagram
has traditionally been one of the chief methods to examine
the stellar population properties of galaxies and their evolution.

The importance of the color-magnitude diagram in galaxy studies can
hardly be overstated.  Very early studies
\citep{devaucouleurs61,visvanathan77} recognized the existence
in such a diagram of a ``red sequence'' (hereafter, RS), where
early-type galaxies lie, also known as the ``color-magnitude relation'' (CMR)
because it represents a positive correlation between galaxy color and
luminosity.

The RS represents an easily recognizable feature in the
color-magnitude diagram that can be used, to first approximation, to
separate red, passively evolving galaxies devoid of star formation,
from bluer star-forming galaxies.
 
Innumerable studies have investigated the origin of such a relation,
and the reasons why a galaxy belongs to it, or falls below it at bluer
colors.

The slope, scatter and location of the RS have long been used to place
constraints on the formation epoch of stars in early-type galaxies
\citep{bower92,ellis97, kodama98}.  The RS is most conspicuous in
galaxy clusters, that are rich in passive early-type galaxies. A
well-defined RS has been observed in clusters up to high redshifts
\citep{lidman08, mei09, strazzullo10} and the fact that the galaxies
with the oldest stars preferentially
inhabit clusters at any epoch can be used to
identify galaxy clusters, using their RS, up to high redshifts
\citep{gladders05, muzzin09, wilson09, gilbank11}.

Observations of 
the relative fractions of galaxies on and below the RS have uncovered
evolution in the star formation activity within clusters \citep{bo84,
ellingson01, loh08} and, more recently, in the field
\citep{bell04,faber07,bell07}.  It has thus become clear that a large
number of galaxies have stopped forming stars and have turned from
blue to red at $z$ below 1 in all environments, but in a way that
depends on environment. It is also now well established that this
progressive ``passivization'' process of galaxies proceeds in a
downsizing fashion, with more massive and luminous galaxies reaching
the RS at earlier epochs than lower-mass, fainter galaxies \citep{cowie96}.
Large spectroscopic surveys have confirmed and placed on a very solid
statistical ground earlier results on the correspondence between the
bimodality in colors (red and blue) and other main galaxy
characteristics such as morphological type (early and late) and galaxy
stellar mass \citep{strateva01,
kauffmann03, cheng10}.

One of the main outstanding questions remains the physical reason why
galaxies stop forming stars and become red. Naturally, there can be
multiple physical mechanisms responsible for the end of star
formation. Their relative importance might vary with galaxy mass,
galaxy location (environment), and redshift, among other things
(Peng et al. 2010).  

Two mechanisms in particular have been the subject of extensive
theoretical investigation in the last years: AGN quenching, and
strangulation by which the hot gas reservoir is removed from galaxies
upon merging with a larger halo \citep{croton06, bower06, vandenbosch08,
guo10, mccarthy10}.
Models however are currently unable to reproduce the observed trends, and 
generally overproduce the fraction of red galaxies
\citep{font08, balogh09, kimm09}. In clusters, additional mechanisms
are expected and observed to take place, such as ram pressure
stripping \citep{gunngott} and harassment \citep{moore98}, though they may be
much more efficient than originally thought even in lower mass systems than
clusters \citep{bekki09}.

The main issue is clearly to what extent the end of star formation is
due to internal galaxy properties, and to what extent it is related
to the global environment (for example the halo mass of the galaxy 
host structure) or the local environment.
In this paper we investigate
the relation between the RS properties
of galaxies in clusters and the global and local environmental
properties in the local Universe. We use a survey of 77
X-ray selected nearby 
galaxy clusters (the WIde-field Nearby Galaxy-cluster Survey, 
WINGS) to search for correlations between the RS
characteristics and the galaxy environment, with the aim of
shedding some light on the mechanisms
that result in the population of passive red galaxies in clusters today. 
WINGS clusters span a wide range of cluster masses, and their galaxies
are found in widely different local environmental densities, thus allowing
a broad investigation using a  homogeneous, high quality photometric dataset.


In the following, we describe the WINGS dataset (\S2) and the method used
to define the RS (\S3). In \S4 we investigate if the RS characteristics
(slope, scatter and value of the
best fit line at $M_V=-20$), the ratio of the number of luminous to faint
galaxies on the RS, the blue galaxy fraction and the morphological
fractions of galaxies on the RS depend on general cluster properties
(cluster velocity dispersion, X-ray luminosity, BCG dominance, number of 
substructures and concentration of ellipticals). We then study
the same characteristics of the RS as a function of local galaxy
density (\S4.5). Our conclusions are summarized in \S5.

Throughout this paper we use the cosmological parameters 
($H_0$, ${\Omega}_m$, ${\Omega}_{\lambda}$) = 
(70 $\rm km \, s^{-1} \, Mpc^{-1}$, 0.3, 0.7).


\section{The Data}\label{sec:data}
The galaxies examined in this paper are part of the \textit{WIde-field
Nearby                     Galaxy-clusters                     Survey}
\citep[][]{wingsI}. WINGS\footnote{Please  refer to WINGS  Website for
updated       details  on    the      survey     and   its   products,
\texttt{http://web.oapd.inaf.it/wings} }  is a  multiwavelength survey
designed  to  provide a robust characterization of
the photometric and  spectroscopic  properties of galaxies   in nearby
clusters,  and to determine  the variations of  these  properties as a
function of galaxy mass and environment.

Clusters  were selected in the  X-ray from the ROSAT Brightest Cluster
Sample and its extension  \citep[][]{ebeling98,ebeling00} and  the X-ray
Brightest  Abell-type Cluster sample    \citep[][]{ebeling96}.   WINGS
clusters cover  a wide range   of velocity dispersion $\sigma_{clus}$,
typically between 500   and   1100 $\rm km  \,  s^{-1}$,    and X-ray
luminosity $L_X$, typically $0.2-5 \times 10^{44} \rm erg/s$.

The survey core dataset, consisting of 
optical  B and V imaging of 78 nearby ($0.04 < z
< 0.07$) galaxy-clusters \citep[][]{wingsII}, has been complemented by
several  ancillary  projects:  (i)  a  spectroscopic follow  up  of 
about 6500 galaxies in a
subsample of  48 clusters, obtained with  the spectrographs WYFFOS@WHT
and 2dF@AAT \citep[][]{wingsspe};  (ii) near-infrared (J, K) imaging of
a    subsample   of    28   clusters    obtained    with   WFCAM@UKIRT
\citep[][]{wingsIII}  ;  (iii)  U  broad- and  H$_\alpha$  narrow-band
imaging of subsamples   of  WINGS clusters, obtained   with wide-field
cameras at     different telescopes \citep[INT,    LBT, Bok,  see][in
preparation]{omizzolo10}.

The B and V photometric catalogs used in this paper are described
in detail in \cite[][]{wingsII}. Briefly, our catalogs are 90\% complete 
at $V \sim 21.7$, and the star-galaxy separation was visually checked,
making the number of misclassifications negligible, down to $V = 22$.
The distance moduli of our clusters are in the range 36.1-37.4, therefore
our photometric catalogs are highly complete and reliable down to $M_V = -15.7$
or fainter.
The photometry was performed on images in which large galaxies and halos 
of bright stars were removed after modeling them with elliptical isophotes.
This procedure greatly improved the photometry of the large galaxies
and increased by 16\% the detection rate of objects projected onto
them  \citep[][]{wingsII}.

WINGS  galaxy morphologies were derived from V  images using the purposely
devised tool  MORPHOT (Fasano et al., 2011 submitted, and Appendix A in 
Fasano et al. 2010).  Our approach
is   a generalization   of   the  non-parametric  method  proposed  by
\citet{conselice00} \citep[see also,][]{conselice03}. In particular, we have
extended the   classical    CAS (Concentration/  Asymmetry/clumpinesS)
parameter set by introducing  a number of additional, suitably devised
morphological  indicators,  using  a final   set of 11
parameters.  A
control sample of 1,000 visually classified  galaxies has been used to
calibrate the whole  set of morphological indicators,  with the aim of
identifying the best  sub-set among them, as  well as of analyzing how
they  depend  on    galaxy  size,  flattening  and   S/N   ratio.  The
morphological  indicators   have been   combined with  two independent
methods, a Maximum Likelihood analysis and a Neural Network trained on
the control   sample of   visually classified  galaxies.    The final,
automatic morphological classification  combines the  results of  both
methods.  We    have  verified  that   our   automatic   morphological
classification reproduces well the  visual classification by two
of us  (AD and GF). In  particular, the robustness and  reliability of
the MORPHOT results turn out to  be comparable with the typical values
obtained comparing the visual classifications  obtained by
different  experienced     human       classifiers      \citep[][in
preparation]{morphot}.      


All the following analysis is carried out for galaxies
within $R_{200}/2$, where $R_{200}$ (usually considered
an approximation for the cluster
virial radius) is the radius enclosing the sphere
with interior mean density 200 times the critical density of the Universe
at that redshift.
The  $R_{200}/2$ values were measured from the cluster velocity dispersions
given in Table 1 as
\begin{equation}
R_{200} = 1.73 \, \frac{\sigma}{1000 \, \rm km \, s^{-1}} \,
{\frac{1}{\sqrt{{\Omega}_{\Lambda} + {\Omega}_{0}(1+z)^3}}} \, h^{-1} \, \rm Mpc\end{equation}
After careful visual inspection of all the B-V versus V
color-magnitude diagrams (CMDs) of our clusters, we decided to include
only 72 clusters in this work because 
of incomplete radial coverage
out to $R_{200}/2$ or nearly absent red sequence
(RS).  

In Table.\ref{tab:stats}, we present a number of global properties of
our clusters. Cluster redshifts 
are from \citet{wingsspe}. 
Cluster velocity dispersions were derived, following the
recipes given in Cava et al. (2009), from the WINGS database that
collects, together with our data, all the available redshifts
from NED.
X-ray total luminosities are taken from Ebeling et al. (1996, 1998, 2000), 
and they have been converted to our adopted cosmology.
The number of substructures found in each cluster are drawn from the
analysis of \citet{ramella07}, who
searched for substructures in the projected distribution of galaxies
in WINGS images  using an adaptive-kernel procedure.
 The BCG prevalence value is determined
with the following equation:
\begin{equation}
\rm{BCGprev}=(V_{\rm{rank2}}+V_{\rm{rank3}})/2-V_{\rm{rank1}}
\end{equation}    
where the V magnitudes refer to the first (the BCG), second and third
ranked galaxies of the cluster. The higher the BCG prevalence value, the
greater is the separation in magnitudes between the BCG and
the next brightest cluster galaxies. It is believed that
a system of galaxies, where most of the
mass has been assembled very early, develops a larger magnitude
gap between the brightest and second brightest galaxies
compared to systems that form later (Dariush et al. 2010 and references 
therein), therefore the BCGprev value would be somehow related to
the main cluster ``assembly epoch''.
The concentration of elliptical galaxies (EsConc) is measured through
a modified version of the Gini coefficient \citep{abra03,lotz04}.  In
particular, 
we define
this coefficient as the difference between the area subtended by
the cumulative distribution function of the elliptical galaxies
(rank-ordered according to their cluster-centric distance) and the
diagonal of the square \citep[Lorentz curve; for a perceptual
illustration see figures~1 in][]{abra03,lotz04}.
In this formulation, the Gini coefficient increases as
the concentration of galaxies toward the cluster center increases.

Finally, we measured galaxy local densities for each galaxy
in our sample using the circular area (A$_{10}$)
containing the 10 nearest projected neighbors in the photometric
catalog (with or without spectroscopic membership) whose V-band
absolute magnitude would be $\leq -19.5$ if they were cluster
members. As we only want to count as neighbours the members of the
cluster, a statistical field correction has been applied to the counts
using Table~5 in \cite{berta06}. In particular, since the field counts
in the area containing the 10 nearest neighbors are not integer
numbers, A$_{10}$ is obtained interpolating the two A$_n$ areas for
which the corrected counts (or the number of spectroscopic members, if
greater than them) are immediately lower and greater than 10.
A similar interpolation technique has also been used when the circular
area containing the 10 nearest neighbors is not fully covered by the
available data (galaxies at the edges of the WINGS field). In this
case, at increasing $n$ (and the corresponding area A$_n$), a coverage
factor has been evaluated as the ratio between the circular area and
the area actually covered by the observations. Then, the counts $n$
have been scaled upwards to account for the corresponding coverage factors and
then have been corrected for the field counts. 
Finally, as in the previous case, A$_{10}$
has been obtained by interpolating the two A$_n$ areas for which the
corrected counts are immediately lower and greater than 10.


\section{Red Sequence and Field Subtraction\label{sec:fieldsub}}

One of the main issues in defining the red sequence of our clusters is
the  removal   of the interloper galaxies. For all
galaxies with  redshift   information, either from  WINGS   spectra or
literature data,  the cluster membership is  known \citep[][]{wingsspe}.
In the  absence of  spectroscopy,   the only alternative way is a statistical approach.

\subsection{The interloper issue}
The Monte Carlo statistical field subtraction tecnique is the standard
method used \citep[see \eg][]{kodama01,pimbblet02}. We used the
\citet{berta06} catalog of field galaxies in the ELAIS-S1 area
(hereafter simply called FIELD), a cluster-free portion
of the sky,
to determine the number of contaminating galaxies as a
function of magnitude and color.  Both the WINGS (which
comprehend cluster+field galaxies) and the FIELD data have been
conveniently binned onto a grid in the color-magnitude diagram space.
Obviously, for each cluster, the numbers of field galaxies in the grid
are rescaled to the size of the field covered by WINGS.

Following
\citet{pimbblet02}, we defined  the probability for  each
galaxy to be an interloper as follows:
\begin{equation}
P_{\rm{int}}(col,mag) \equiv
\frac{N_{\rm{F}}(col,mag)}{N_{\rm{W}}(col,mag)} \label{eq:pfield}
\end{equation}
where $N_{\rm{F}}$ and  $N_{\rm{W}}$  are the number of FIELD and WINGS
galaxies in each $(col,mag)$ bin, respectively. We further expanded
the method to include all the redshift
information we have: 
spectroscopically
confirmed members $N_{\rm{YES}}$ and non-members $N_{\rm{NO}}$
were always retained and rejected, respectively, in the analysis. 

Eq.(\ref{eq:pfield}) was modified accordingly:
\begin{equation}
P_{\rm{int}} \equiv
\frac{N_{\rm{F}}(col,mag)-N_{\rm{NO}}(col,mag)}{N_{\rm{W}}(col,mag)-N_{\rm{NO}}(col,mag)-N_{\rm{YES}}(col,mag)} \label{eq:pfield1}
\end{equation}

One of the main caveats of this method is the possibility that the
probability becomes negative: in this case we used the same recipe of
\citet{pimbblet02}, by applying an adaptive grid tecnique
\citep[for further details see][Appendix A]{pimbblet02}. Once the
interloper
probability was determined at each grid position, 
we performed 100 Monte Carlo simulations
generating a number between $0.0$ and $1.0$ for each galaxy.  If the
random number was greater than the corresponding $P_{\rm{int}}$, than
the galaxy was flagged as a member, otherwise it was assigned to the
interloper population.

With this procedure the assumption that the FIELD sample is a {\it perfect} realization of the {\it real} interloper population is mitigated. Indeed, it can be
demonstrated that with this procedure we are actually giving
to each bin of the FIELD with $N$ galaxies, a corresponding poissonian
error of $\sqrt{N}$; thus, it  is like extracting at each run a
new FIELD sample (a sort of bootstrap).

\begin{figure*}
\centering
\includegraphics[scale=0.7]{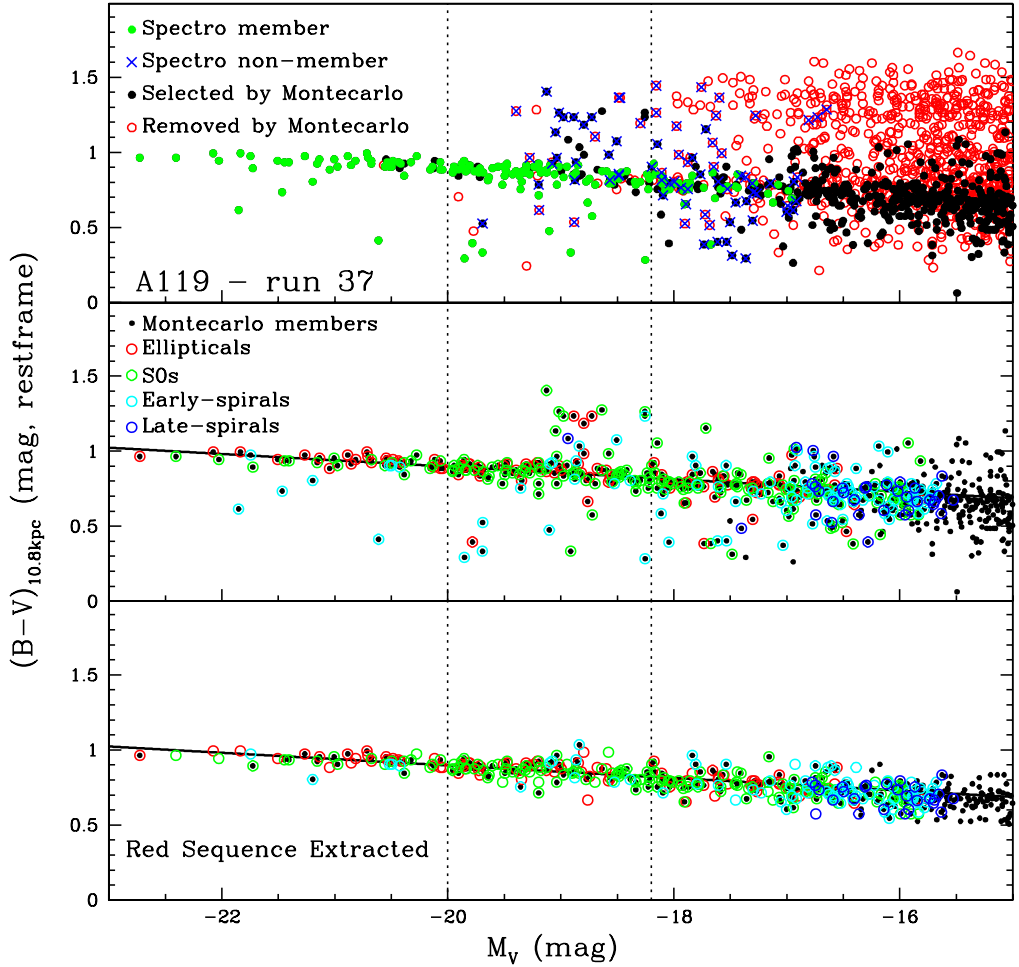}
\caption{Visual example of a Monte Carlo realization for the cluster
Abell 119. In the upper panel, the red empty dots and the black filled
dots are the non-members and members assigned by the simulation (run
37th), respectively. Shown as green circles and blue crosses are the
spectroscopic members and non-members, respectively. In the middle
panel only members according to the Montecarlo realization
are shown, but divided in different morphological
types (see legend and the figure in electronic format). In the bottom
panel, the RS extracted is shown, again with the morphological color
coding. The dotted vertical lines mark the absolute V magnitude
limits used to define the Luminous and Faint population on the
RS.
\label{fig:bgsub}}
\end{figure*}

In Fig.\ref{fig:bgsub} we show the 37-th run of a Monte Carlo
realization of the cluster Abell 119. In the upper panel all galaxies
from the catalog are presented, but shown as green circles and blue
crosses are the spectroscopic members and non-members,
respectively. The use of the spectroscopic membership information is the
only difference from the interloper subtraction used in
\citet{pimbblet02}. Red open circles and black filled dots refer to
the galaxies assigned to the non-member and member population by the
simulation. We also show the different morphologies, when available
(see legend in the figure), of all cluster members (central panel) and
only RS galaxies, as defined in the next section 
(bottom panel). Note the non-negligible presence of
red spirals (blue and cyan open circles).

\subsection{Defining the red sequence} \label{sec:rs}
We calculate rest frame absolute magnitudes and colors by applying a 
galaxy-type dependent
k-correction based on the observed (B-V) color of the galaxies within
an aperture (diameter) of 10.8kpc using the K-corrections
from \cite[][]{poggianti97}.
The physical aperture assures a well-sampled,
distance-independent color for all galaxies. 
We further apply a cut in color
$0.5\leq(B-V)_{10.8kpc}\leq1.5$ to eliminate those eventual (few)
galaxies that
are surely either too red or too blue to belong to our clusters but might have
remained after the statistical subtraction. 
In the following, we consider only galaxies brighter than 
$M_V\sim-18$ for which both our photometric and morphological
catalogs are complete (Varela et al. 2009, Fasano et al. 2011).

We use the robust bi-weight estimator fitting procedure in two steps
to identify the RS. The {\em first step} consists of fitting all the
galaxies from the previous selection process with absolute V magnitude
$-21.5\leq M_V\leq -18$ and calculating the median distance $d_{red}$
of all the galaxies from the best fit. We choose this magnitude 
interval because
most of our CMDs (see Fig.\ref{fig:bgsub}) exhibit a change of slope
(are shallower) at 
higher luminosities, .

In the {\em second step} only galaxies with $-21.5\leq M_V\leq -18$
and distance $3\times d_{red}$ from the previous step are fitted, and
the RMS of the residuals is calculated. Only galaxies within
0.2 magnitudes
of the second fit are assigned to the
RS-population.\footnote{Using galaxies within $2\times \rm{RMS}$ from the fit 
did not change any of the results of this paper.}
All non RS-galaxies bluer than that are
assigned to the blue-population.

This procedure is applied to each of the 100 Montecarlo Simulations and a final
mean value and its error are found for each individual
cluster (see Table\ref{tab:stats}). In Fig.(\ref{fig:cmd1})
we present a sample plot of one 
of the Montecarlo realizations for each cluster, with the best fit
parameters and the global scatter for the red sequence and the main
properties of the cluster shown at the bottom of each panel. The vertical lines
indicate the V-band absolute magnitude limits used to define the ``Luminous''
and the ``Faint'' red sequence population (see section
\ref{sec:lumtofaint}).

\begin{figure*}
\centering
\includegraphics[scale=0.8]{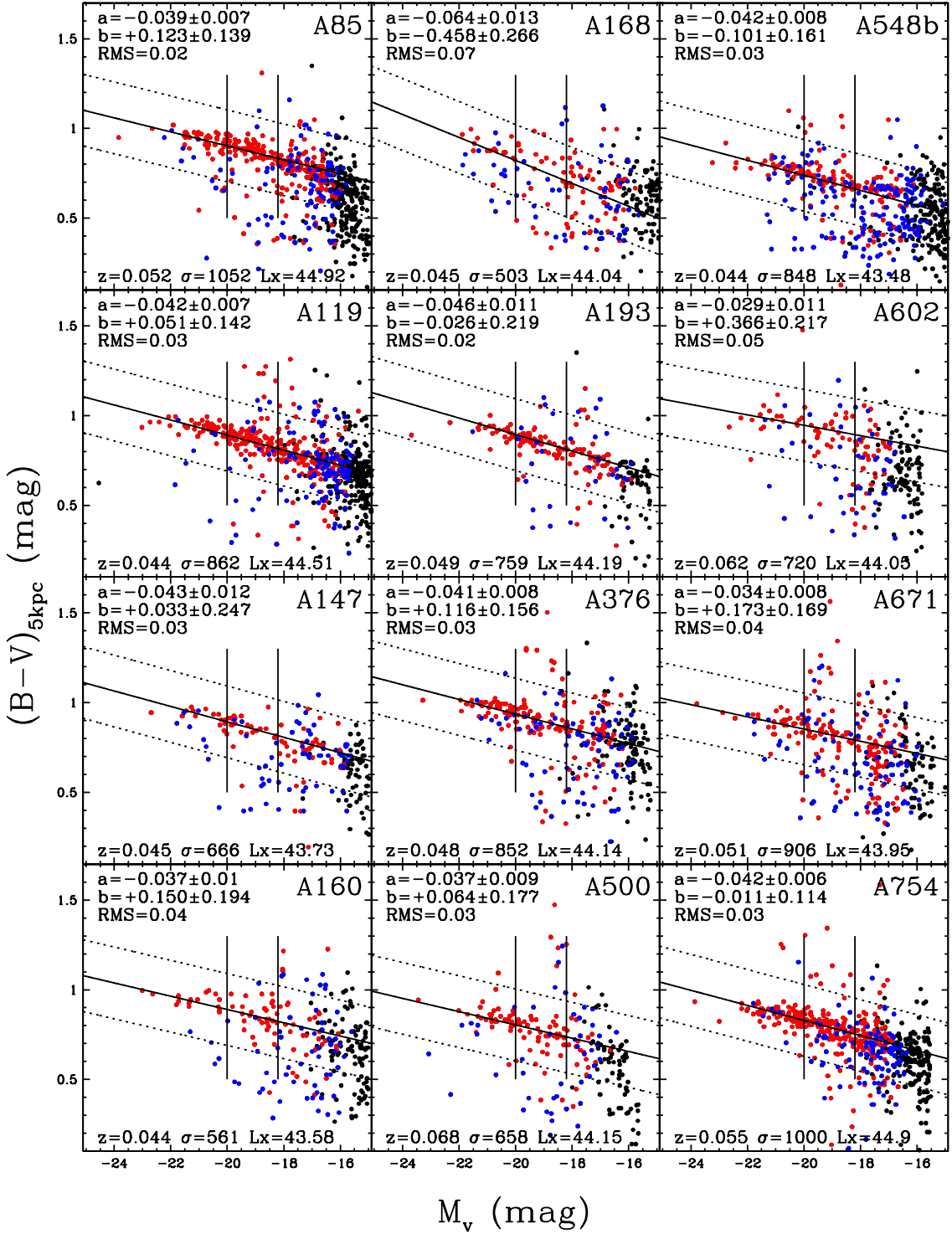}
\caption{
A random realization of the Montecarlo interloper subtraction
technique. Only galaxies assigned to the clusters are shown: red dots
are morphologically
early-type galaxies (Es and S0s), blue dots are late-type ones and black dots
are galaxies with no available morphological classification (colors
only available in the on-line version). The best
fitted RS of the 100 realizations is drawn (solid black line) with
$\pm0.2$mag limits (dashed black lines) identifying the red sequence.
The RMS numerical values are reported at the top of each panel. The global
properties of each cluster are reported, for reference, at the bottom
of each panel. The vertical lines mark the absolute V magnitude limits
used to define the Luminous and the Faint population of the RS.
\label{fig:cmd1}}
\end{figure*}

\addtocounter{figure}{-1}
\begin{figure*}
\centering
\includegraphics[scale=0.9]{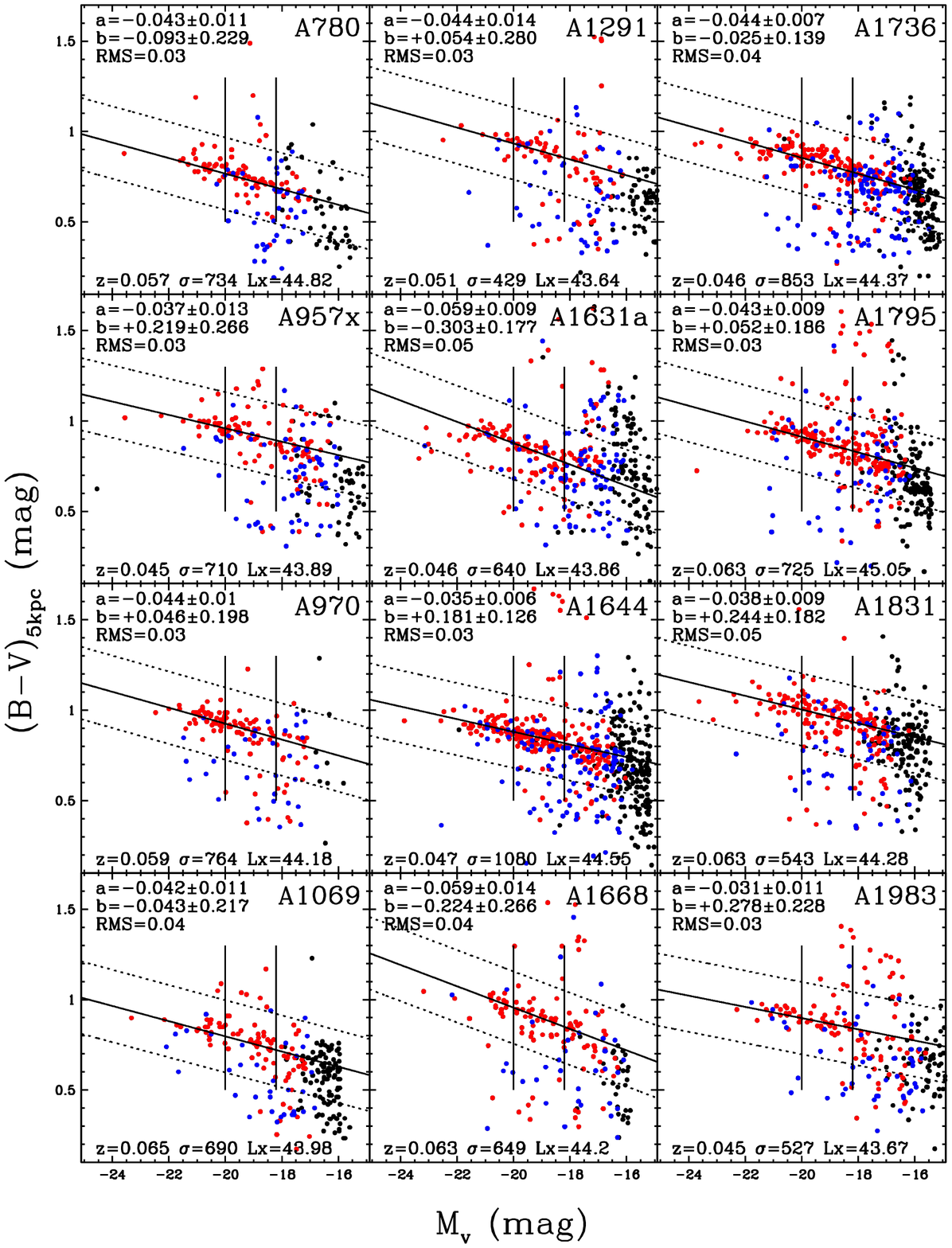}
\caption{continued.
\label{fig:cmd2}}
\end{figure*}

\addtocounter{figure}{-1}
\begin{figure*}
\centering
\includegraphics[scale=0.9]{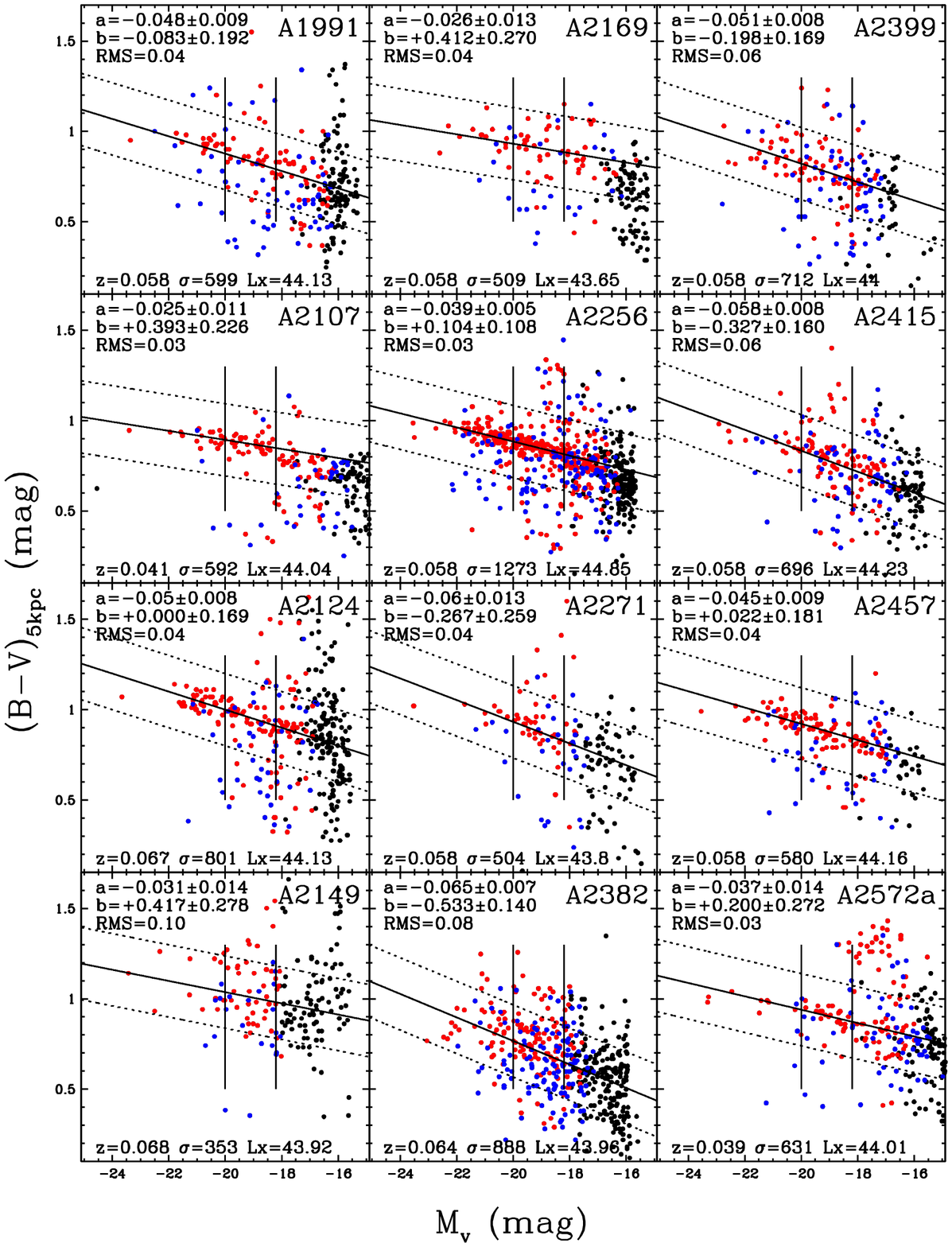}
\caption{continued.
\label{fig:cmd3}}
\end{figure*}

\addtocounter{figure}{-1}
\begin{figure*}
\centering
\includegraphics[scale=0.9]{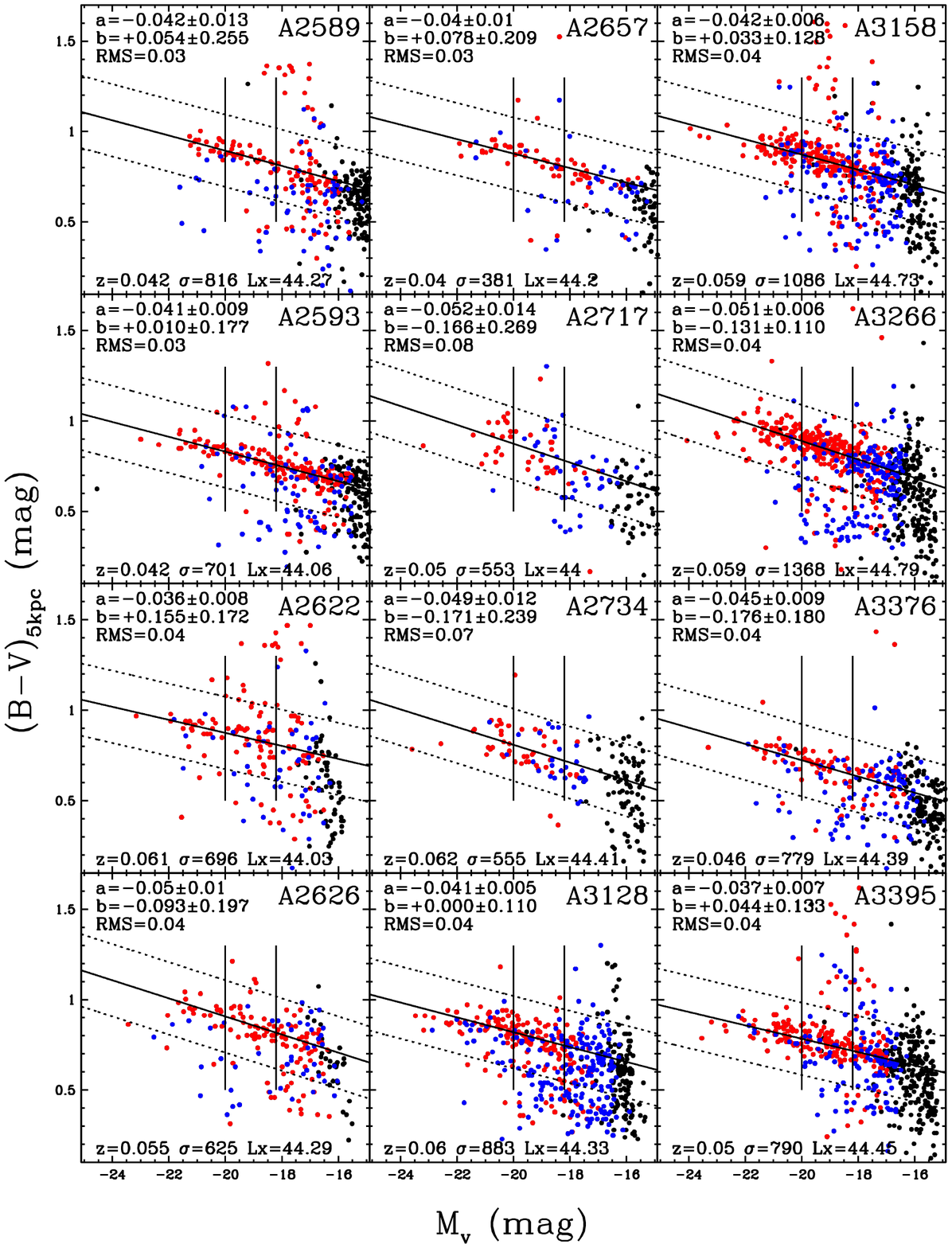}
\caption{continued.
\label{fig:cmd4}}
\end{figure*}

\addtocounter{figure}{-1}
\begin{figure*}
\centering
\includegraphics[scale=0.9]{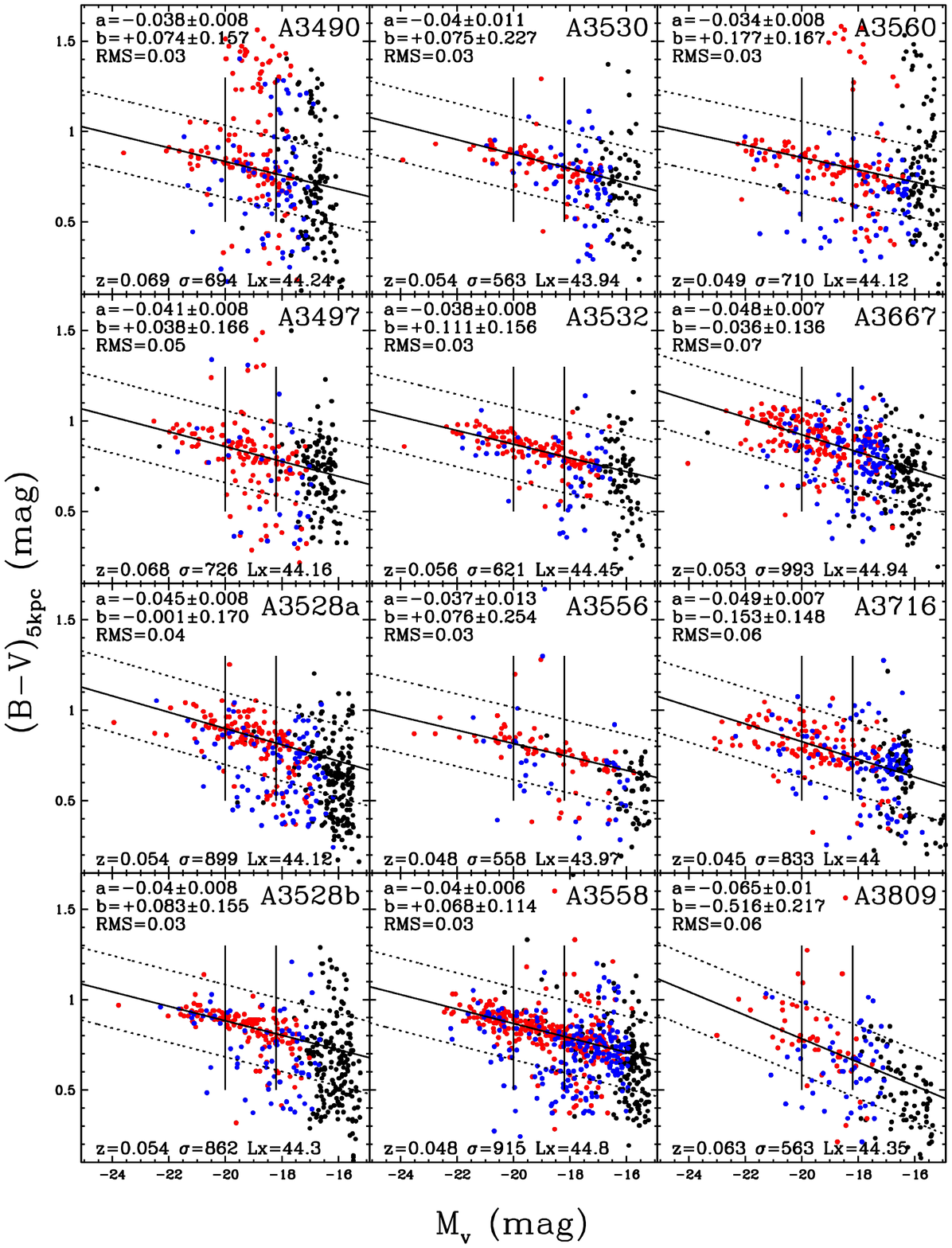}
\caption{continued.
\label{fig:cmd5}}
\end{figure*}

\addtocounter{figure}{-1}
\begin{figure*}
\centering
\includegraphics[scale=0.9]{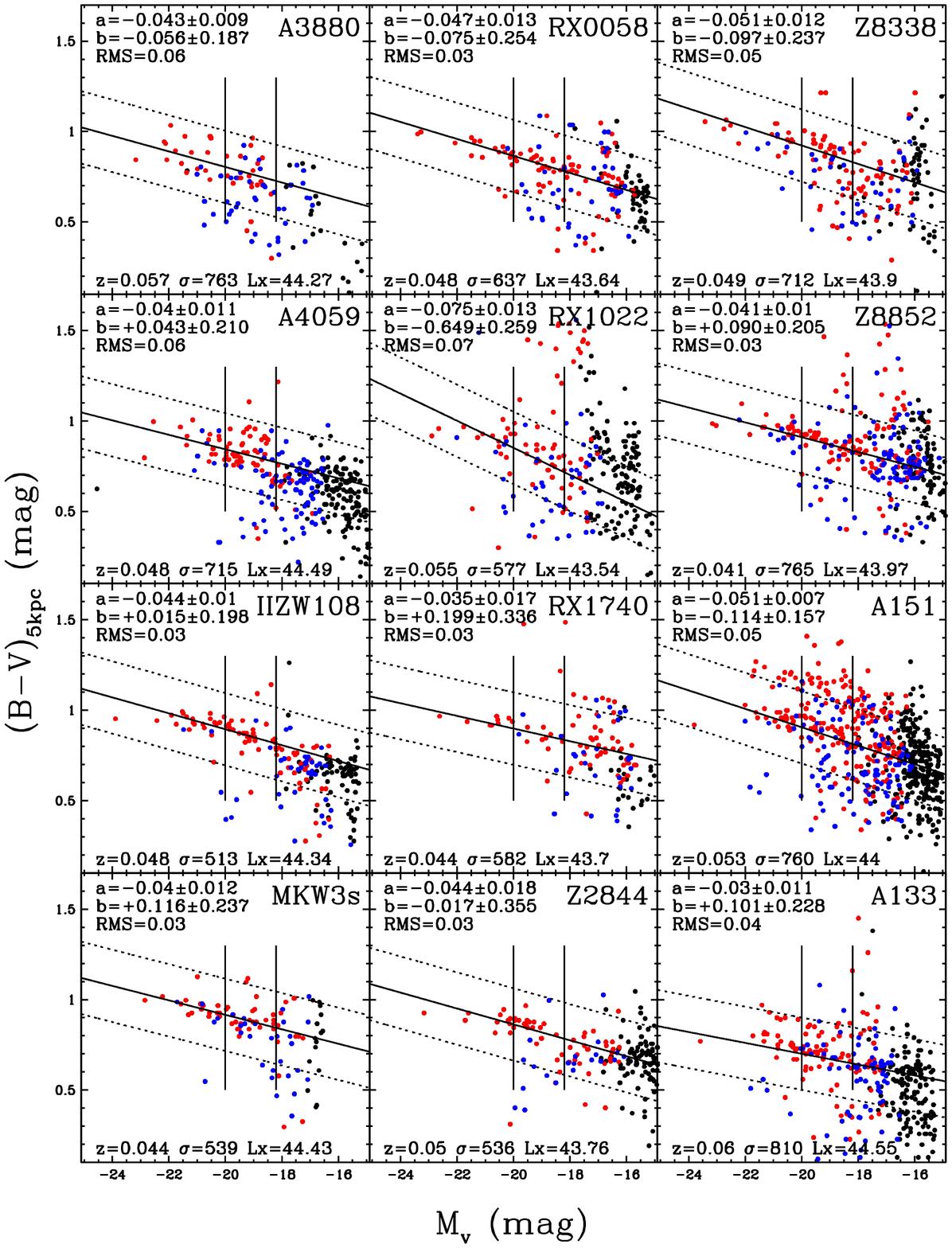}
\caption{End of Figure \ref{fig:cmd1}.
\label{fig:cmd6}}
\end{figure*}

\begin{figure*}
\centering
\includegraphics[scale=0.7]{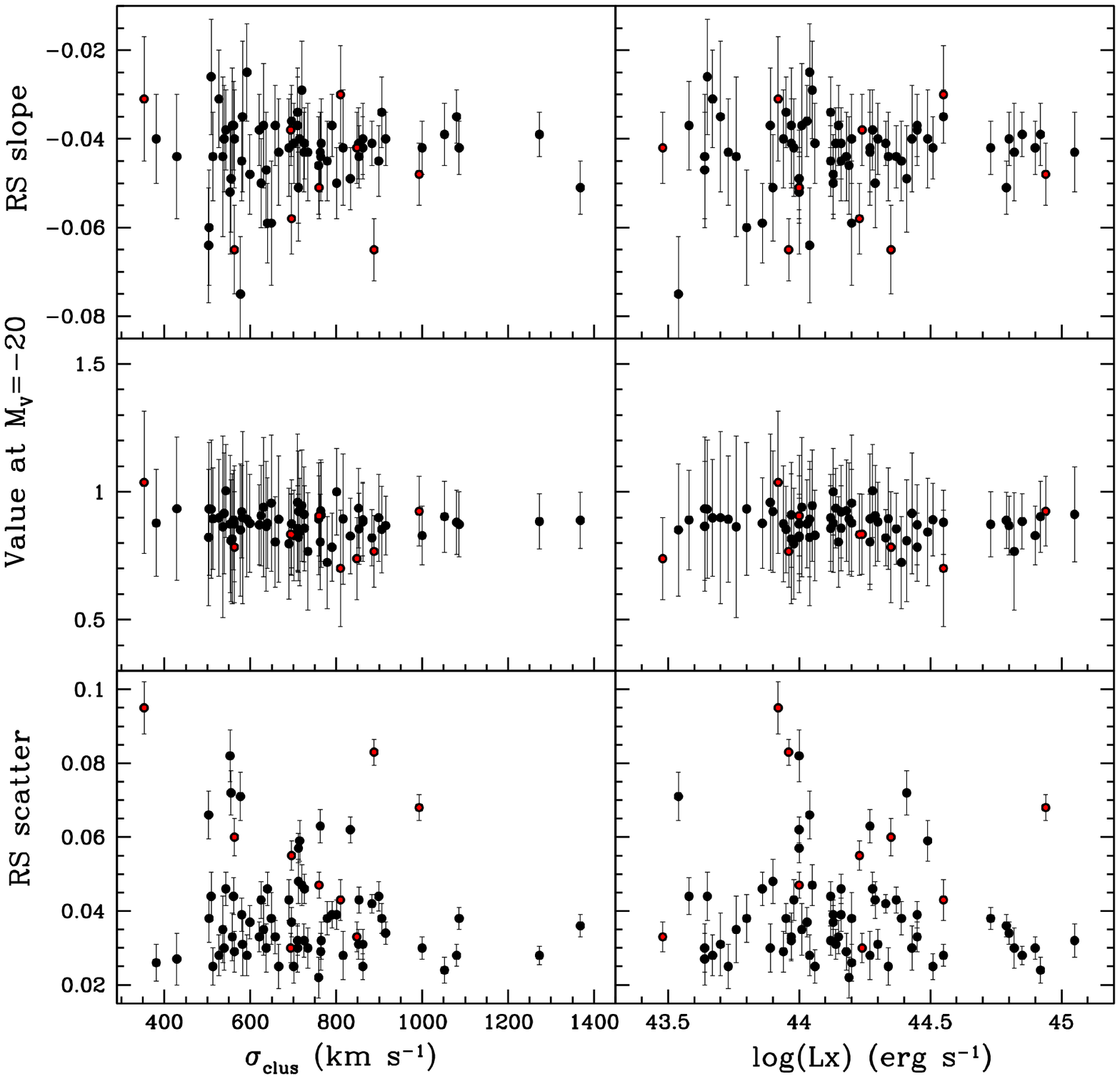}
\caption{Slope, value of best fit line at $M_V = -20$
and scatter of the median RS values of the 100 Montecarlo runs for each cluster versus cluster velocity dispersion ($\sigma_{clus}$, left panels) and cluster X-ray luminosity ($L_X$, right panels). Error bars refer to the error of the median of the 100 simulations. In red (on-line version),
clusters whose RS is contaminated by another structure (see text).
\label{fig:cmddep}}
\end{figure*}

\section{Results}

\subsection{Slope and scatter}

In Table\ref{tab:stats}, in addition to the global parameters of
our cluster sample, we present the median slope and scatter of the 100
Monte Carlo realizations described in section \ref{sec:rs}.  The
median slope for all our clusters is $-0.047\pm0.001$, and the typical
scatter is of the order of $\sim0.05$mag.

In Fig.\ref{fig:cmddep} we plot the slope, the fit value at $M_V=-20$
and the scatter of the RS against the cluster velocity dispersion and
total X-ray luminosity (both of which are related to the cluster
mass). We have coloured in red those clusters whose RS appears to be
contaminated by another galaxy structure visible in Fig.~2 (A133,
A151, A548b, A2149, A2382, A2415, A3490, A3667, A3809),\footnote{We
have not highlighted here those clusters with a second RS that lies at
sufficiently higher redshift not to be expected to influence the RS of
the main WINGS structure, such as A2572, A3560, A3490, RX1022.}
although it should be kept in mind that several other clusters have a
very broad RS suggestive of more than one structure along the line of
sight and close in redshift to the main cluster.

If the epoch and rate at which galaxies turn red and
reach the color-magnitude RS
depended upon the 
mass of the cluster to which galaxies belong at $z=0$,
we should observe a correlation between cluster mass
and RS properties.
Instead, we do not find any such correlation
in our sample of 72 local clusters. 

Similarly, we do not find any dependence of the RS
slope, scatter, and $M_V=-20$ value on the number of substructures,
the BCG prevalence value and the concentration of ellipticals
(not shown).
Our results agree with the previous findings of \citet{stott09}
which showed a lack of correlation between the RS slope and the cluster
$\sigma$, $L_X$ and BCG dominance at $z=0.1-1$.


\begin{table*}
\begin{center}
\caption{ \label{tab:stats}}
\begin{tabular}{lcccccccc}
Cluster & z & $\sigma_{\rm{clus}}$ & $\log(\rm{Lx})$ & Nsub & BCGprev & EsConc & slope & RMS \\
\hline
\hline
A85 & 0.052 & 1052.0 & 44.92 & 2 & 1.405 & 0.236 & $-0.039\pm0.007$  &  0.024 \\
A119 & 0.044 & 862.0 & 44.51 & 2 & 0.990 & 0.179 & $-0.042\pm0.007$  &  0.025 \\
A133 & 0.060 & 810.0 & 44.55 & -- & 0.575 & 0.067 & $-0.030\pm0.011$ &  0.043 \\
A147 & 0.045 & 666.0 & 43.73 & 2 & 0.505 & 0.162 & $-0.043\pm0.012$  &  0.025 \\
A151 & 0.053 & 760.0 & 44.00 & 2 & 1.220 & 0.108 & $-0.051\pm0.007$  &  0.047 \\
A160 & 0.044 & 561.0 & 43.58 & 3 & 0.485 & 0.119 & $-0.037\pm0.010$  &  0.044 \\
A168 & 0.045 & 503.0 & 44.04 & 1 & 1.130 & 0.128 & $-0.064\pm0.013$  &  0.066 \\
A193 & 0.049 & 759.0 & 44.19 & 0 & 1.000 & 0.164 & $-0.046\pm0.011$  &  0.022 \\
A376 & 0.048 & 852.0 & 44.14 & 1 & 1.345 & 0.165 & $-0.041\pm0.008$  &  0.031 \\
A500 & 0.068 & 658.0 & 44.15 & 1 & 0.435 & 0.175 & $-0.037\pm0.009$  &  0.033 \\
A548b & 0.044 & 848.0 & 43.48 & -- & 0.635 & 0.127 & $-0.042\pm0.008$&  0.033 \\
A602 & 0.062 & 720.0 & 44.05 & 1 & 0.595 & 0.118 & $-0.029\pm0.011$  &  0.047 \\
A671 & 0.051 & 906.0 & 43.95 & 2 & 1.070 & 0.207 & $-0.034\pm0.008$  &  0.038 \\
A754 & 0.055 & 1000.0 & 44.90 & 2 & 1.140 & 0.164 & $-0.042\pm0.006$ &  0.030 \\
A780 & 0.057 & 734.0 & 44.82 & -- & 1.220 & 0.111 & $-0.043\pm0.011$ &  0.030 \\
A957x & 0.045 & 710.0 & 43.89 & 0 & 1.680 & 0.160 & $-0.037\pm0.013$ &  0.030 \\
A970 & 0.059 & 764.0 & 44.18 & 1 & 0.290 & 0.214 & $-0.044\pm0.010$  &  0.029 \\
A1069 & 0.065 & 690.0 & 43.98 & 1 & 1.030 & 0.186 & $-0.042\pm0.011$ &  0.043 \\
A1291 & 0.051 & 429.0 & 43.64 & 1 & 0.785 & 0.144 & $-0.044\pm0.014$ &  0.027 \\
A1631a & 0.046 & 640.0 & 43.86 & 0 & 1.385 & 0.098 & $-0.059\pm0.009$&  0.046 \\
A1644 & 0.047 & 1080.0 & 44.55 & -- & 1.275 &0.000 & $-0.035\pm0.006$&  0.028 \\
A1668 & 0.063 & 649.0 & 44.20 & -- & 1.060 & 0.153 & $-0.059\pm0.014$&  0.038 \\
A1736 & 0.046 & 853.0 & 44.37 & 5 & 0.145 & 0.177 & $-0.044\pm0.007$ &  0.043 \\
A1795 & 0.063 & 725.0 & 45.05 & 1 & 1.480 & 0.185 & $-0.043\pm0.009$ &  0.032 \\
A1831 & 0.063 & 543.0 & 44.28 & 1 & 0.875 & 0.067 & $-0.038\pm0.009$ &  0.046 \\
A1983 & 0.045 & 527.0 & 43.67 & -- & 0.275 & 0.126 & $-0.031\pm0.011$&  0.028 \\
A1991 & 0.058 & 599.0 & 44.13 & 1 & 0.855 & 0.102 & $-0.048\pm0.009$ &  0.037 \\
A2107 & 0.041 & 592.0 & 44.04 & 0 & 1.395 & 0.080 & $-0.025\pm0.011$ &  0.028 \\
A2124 & 0.067 & 801.0 & 44.13 & 0 & 1.945 & 0.124 & $-0.050\pm0.008$ &  0.039 \\
A2149 & 0.068 & 353.0 & 43.92 & 0 & 0.945 & 0.128 & $-0.031\pm0.014$ &  0.095 \\
A2169 & 0.058 & 509.0 & 43.65 & 0 & 0.315 & 0.011 & $-0.026\pm0.013$ &  0.044 \\
A2256 & 0.058 & 1273.0 & 44.85 & 0 & 0.430 & 0.141 & $-0.039\pm0.005$&  0.028 \\
A2271 & 0.058 & 504.0 & 43.80 & -- & 1.260 & 0.102 & $-0.060\pm0.013$&  0.038 \\
A2382 & 0.064 & 888.0 & 43.96 & -- & 0.295 & 0.124 & $-0.065\pm0.007$&  0.083 \\
A2399 & 0.058 & 712.0 & 44.00 & 0 & 0.125 & 0.202 & $-0.051\pm0.008$ &  0.057 \\
A2415 & 0.058 & 696.0 & 44.23 & 1 & 0.380 & 0.159 & $-0.058\pm0.008$ &  0.055 \\
A2457 & 0.058 & 580.0 & 44.16 & 1 & 1.230 & 0.116 & $-0.045\pm0.009$ &  0.039 \\
A2572a & 0.039 & 631.0 & 44.01 & 3 & 0.440 & 0.094 & $-0.037\pm0.014$&  0.035 \\
A2589 & 0.042 & 816.0 & 44.27 & -- & 2.030 & 0.104 & $-0.042\pm0.013$&  0.028 \\
A2593 & 0.042 & 701.0 & 44.06 & 1 & 0.995 & 0.212 & $-0.041\pm0.009$ &  0.025 \\
A2622 & 0.061 & 696.0 & 44.03 & 1 & 1.220 & 0.128 & $-0.036\pm0.008$ &  0.037 \\
A2626 & 0.055 & 625.0 & 44.29 & -- & 1.120 & 0.142 & $-0.050\pm0.010$&  0.043 \\
A2657 & 0.040 & 381.0 & 44.20 & 1 & 0.865 & 0.008 & $-0.040\pm0.010$ &  0.026 \\
A2717 & 0.050 & 553.0 & 44.00 & -- & 1.030 & 0.129 & $-0.052\pm0.014$&  0.082 \\
A2734 & 0.062 & 555.0 & 44.41 & 3 & 0.860 & 0.149 & $-0.049\pm0.012$ &  0.072 \\
A3128 & 0.060 & 883.0 & 44.33 & 3 & 0.315 & 0.114 & $-0.041\pm0.005$ &  0.042 \\
A3158 & 0.059 & 1086.0 & 44.73 & 1 & 0.495 & 0.217 & $-0.042\pm0.006$&  0.038 \\
A3266 & 0.059 & 1368.0 & 44.79 & 0 & 1.065 & 0.235 & $-0.051\pm0.006$&  0.036 \\
A3376 & 0.046 & 779.0 & 44.39 & 2 & 0.870 & 0.089 & $-0.045\pm0.009$ &  0.038 \\
A3395 & 0.050 & 790.0 & 44.45 & -- & 0.575 & 0.138 & $-0.037\pm0.007$&  0.039 \\
A3490 & 0.069 & 694.0 & 44.24 & -- & 0.375 & 0.111 & $-0.038\pm0.008$&  0.030 \\
A3497 & 0.068 & 726.0 & 44.16 & -- & 0.340 & 0.076 & $-0.041\pm0.008$&  0.046 \\
A3528a & 0.054 & 899.0 & 44.12 & -- & 1.470 & 0.135 &$-0.045\pm0.008$&  0.044 \\
A3528b & 0.054 & 862.0 & 44.30 & 1 & 1.595 & 0.216 & $-0.040\pm0.008$&  0.031 \\
A3530 & 0.054 & 563.0 & 43.94 & 0 & 1.310 & 0.111 & $-0.040\pm0.011$ &  0.029 \\
A3532 & 0.056 & 621.0 & 44.45 & 0 & 1.250 & 0.162 & $-0.038\pm0.008$ &  0.033 \\
A3556 & 0.048 & 558.0 & 43.97 & -- & 0.160 & 0.132 & $-0.037\pm0.013$&  0.033 \\
A3558 & 0.048 & 915.0 & 44.80 & 0 & 1.575 & 0.225 & $-0.040\pm0.006$ &  0.034 \\
A3560 & 0.049 & 710.0 & 44.12 & -- & 2.255 & 0.183 & $-0.034\pm0.008$&  0.032 \\
A3667 & 0.053 & 993.0 & 44.94 & 3 & 0.950 & 0.143 & $-0.048\pm0.007$ &  0.068 \\
A3716 & 0.045 & 833.0 & 44.00 & 1 & 0.125 & 0.236 & $-0.049\pm0.007$ &  0.062 \\
A3809 & 0.063 & 563.0 & 44.35 & -- & 1.170 & 0.045 & $-0.065\pm0.010$&  0.060 \\
A3880 & 0.057 & 763.0 & 44.27 & 0 & 0.540 & 0.085 & $-0.043\pm0.009$ &  0.063 \\
A4059 & 0.048 & 715.0 & 44.49 & -- & 1.125 & 0.189 & $-0.040\pm0.011$&  0.059 \\
IIZW108 & 0.048 & 513.0 & 44.34 & 1 & 0.905 & 0.031 &$-0.044\pm0.010$&  0.025 \\
MKW3s & 0.044 & 539.0 & 44.43 & 1 & 0.855 & 0.109 & $-0.040\pm0.012$ &  0.030 \\
RX0058 & 0.048 & 637.0 & 43.64 & 1 & 0.220 & 0.097 & $-0.047\pm0.013$&  0.030 \\
RX1022 & 0.055 & 577.0 & 43.54 & -- & 0.135 & 0.137 &$-0.075\pm0.013$&  0.071 \\
RX1740 & 0.044 & 582.0 & 43.70 & 3 & 0.825 & 0.117 & $-0.035\pm0.017$&  0.031 \\
Z2844 & 0.050 & 536.0 & 43.76 & 2 & 0.865 & 0.038 & $-0.044\pm0.018$ &  0.035 \\
Z8338 & 0.049 & 712.0 & 43.90 & 2 & 0.710 & 0.142 & $-0.051\pm0.012$ &  0.048 \\
Z8852 & 0.041 & 765.0 & 43.97 & 2 & 0.485 & 0.158 & $-0.041\pm0.010$ &  0.032 \\
\hline
\hline
\end{tabular}

\end{center}
\end{table*}

\subsection{Luminous-to-faint ratio and blue fractions} \label{sec:lumtofaint}

The number ratio of luminous-to-faint RS galaxies is often used to investigate 
the evolution of the faint end of the red luminosity function of
cluster galaxies. High-z studies find a deficiency of faint red
galaxies in comparison to local galaxy clusters, as 
the luminous-to-faint ratio is seen to evolve from higher values
at high-z 
to lower values at low-z (see, among others,
\cite{delucia04,tanaka05,delucia07,stott07,gilbank08,lu09,rudnick09,
stott09,capozzi10}
and \cite{andreon08,crawford09} for opposite 
conclusions). 
This is generally
interpreted as evidence for a large number of relatively faint
galaxies having moved to the RS only recently, most probably due to
the quenching of the star formation caused by the high-density
environment. 
The luminous-to-faint ratio has been found to evolve in the field too,
but to be lower in clusters than in the field at all redshifts out to
$z=1$,
implying that the faint end of the red sequence was established first in 
clusters \citep{gilbank08b}.

We wish to study how the luminous-to-faint RS ratio
changes from cluster to cluster and depends on cluster properties
at low redshift.
To define the two galaxy populations 
we corrected our absolute magnitudes for passive evolution:
 we consider all galaxies
brighter than $M_{V}=-20-\rm{lum_{corr}}$ to be \emph{luminous}, 
and all
galaxies fainter than this magnitude and brighter than
$M_{V}=-18.2-\rm{lum_{corr}}$ to be \emph{faint}, 
where $\rm{lum_{corr}}$ is the passive
evolution correction with respect to $z=0$. 
We recall here that these are the same absolute V magnitude limits adopted
by \citet{delucia07} and most other works at high redshift, but that
high-z studies usually consider the (U-V) rest-frame color,
therefore
a direct comparison cannot be performed.

\begin{figure*}
\centering
\includegraphics[scale=0.65,angle=270]{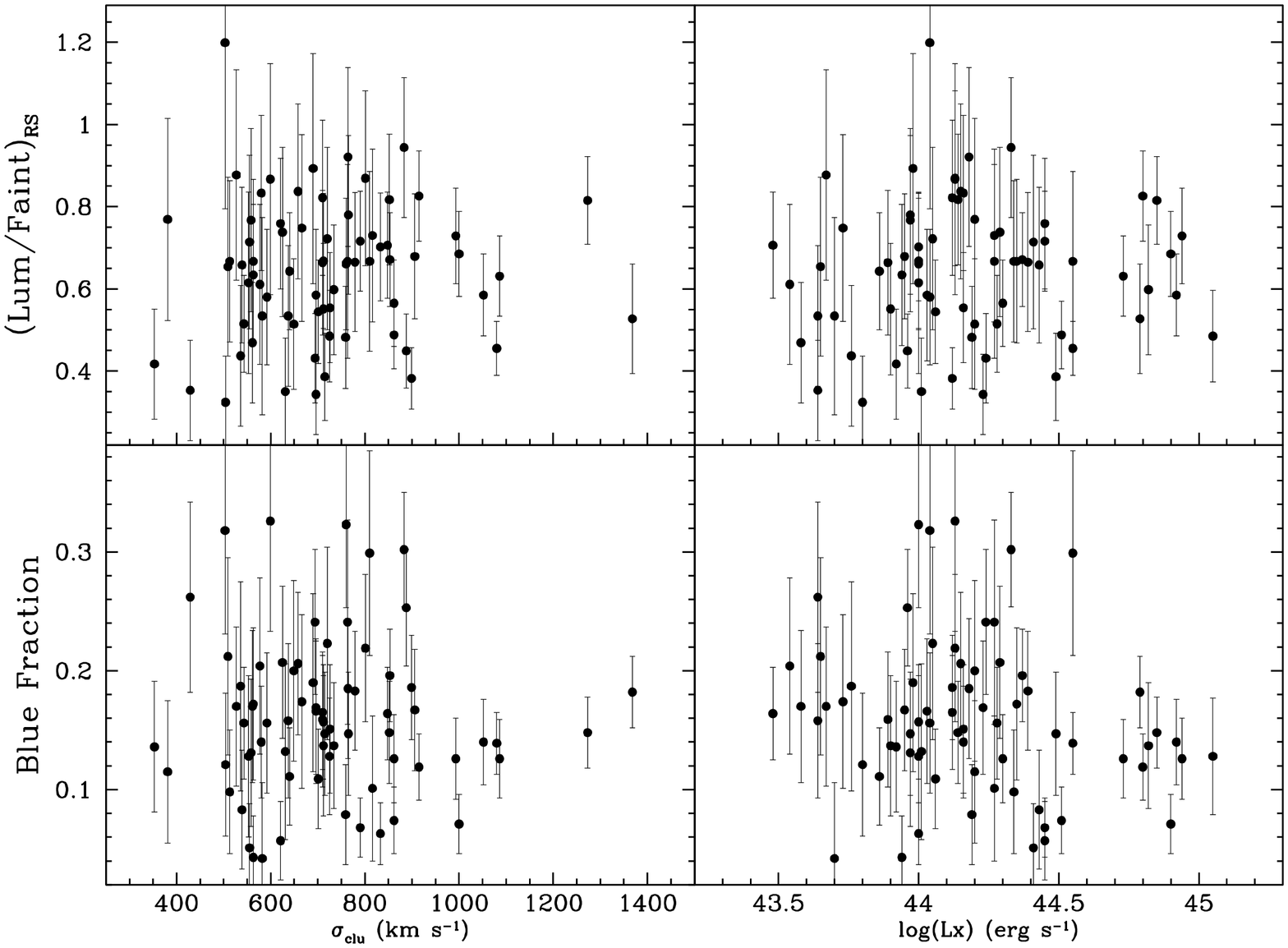}
\caption{Luminous-to-faint ratio (top panels) and blue fraction (bottom panels) vs. cluster velocity dispersion and total X-ray luminosity in WINGS clusters. Errorbars are the mean error on the 100 Montecarlo realizations.
\label{fig:ltofb}}
\end{figure*}

\begin{figure*}
\centering
\includegraphics[scale=0.65,angle=270]{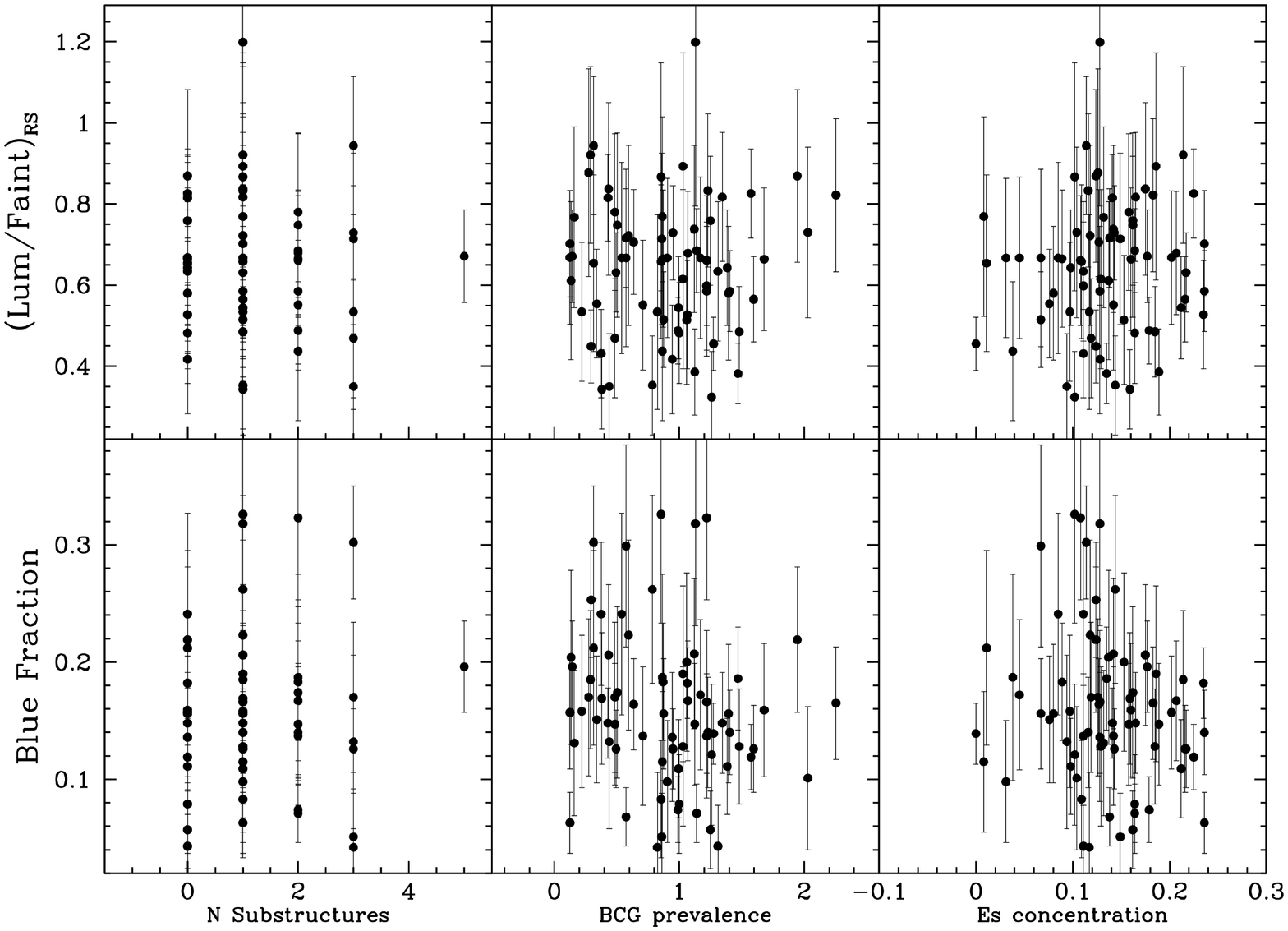}
\caption{Luminous-to-faint ratio (top panels) and blue fraction (bottom panels) vs. number of substructures, BCG prevalence value and concentration of elliptical galaxies in WINGS clusters. Errorbars are the mean error on the 100 Montecarlo realizations.
\label{fig:ltofb2}}
\end{figure*}

Our large sample of local clusters show a large
scatter in the luminous-to-faint ratio.
In the top panels
of Fig.\ref{fig:ltofb} and Fig.\ref{fig:ltofb2} we plot the luminous-to-faint ratio as a
function of cluster velocity dispersion, X-ray
luminosity, number of substructures, BCG prevalence value and concentration
of ellipticals.

As was the case for the RS parameters, no correlation between 
the luminous-to-faint ratio and the cluster properties is found.
This agrees with recent results from \citet{capozzi10} who found
no relationship between luminous-to-faint ratio and cluster
X-ray luminosity at low redshift, and with \citet{rudnick09}
who failed to detect a dependence of the red galaxy luminosity function
on cluster velocity dispersion in the SDSS.

Such lack of a correlation suggests that the downsizing trend in star
formation (the different distribution of
times at which galaxies turn red and reach the RS depending on
their luminosity) does not depend in a simple way from global cluster
properties such as the cluster mass, or level of substructure.
However, the large 
scatter of luminous-to-faint values in our sample
shows that there is a great diversity in the way the RS is populated as a
function of galaxy magnitude. Most likely, this
indicates widely different 
quenching histories in faint galaxies from
cluster to cluster, even for clusters of the same mass at $z=0$. 


The blue fraction is defined as the ratio between the number of blue
galaxies and the total number of galaxies in the cluster, 
for $M_V< -18.2$. In the last
decade the blue fraction has been observed to have a large scatter of
values at any given redshift, and to depend on several obvious parameters 
such as the magnitude limit used and the cluster-centric
distance.  However, contrasting conclusions have been reached as to whether
it depends on cluster properties such as
concentration, richness and
presence of substructure \citep{bo84, wang97, smail98, metevier00, 
ellingson01, margoniner01, pimbblet02, fairley02, balogh04,
depropris04, barkhouse09}. 
All these findings together
imply that environmental effects may compete with, and possibly mimic,
evolutionary trends. Thus, it is of paramount importance to try to
disentangle the cluster-to-cluster variance and the
dependence on cluster properties in order to properly understand and
interpret the studies of high-z clusters and to investigate how
the current galaxy populations in clusters came to be.


Figs.\ref{fig:ltofb} (bottom panels) shows that in WINGS there is no dependence
of the blue fraction on either cluster
velocity dispersion or total X-ray luminosity, as found previously
in the local Universe and at higher redshifts  \citep{fairley02, depropris04,
goto05}, and in agreement with a flat median
star-forming (emission-line) fraction 
for clusters with $\sigma>500 \rm \, km \, s^{-1}$ in
the local Universe  \citep{poggianti06,popesso07}. 
Furthermore, we find no trend with the number of substructures, the
BCG prevalence value and the concentration of elliptical galaxies 
(Fig.\ref{fig:ltofb2}).
A quite large scatter in the blue fraction is observed fixing any of these
parameters, though most clusters have a blue fraction below 20\%, with
a median of 0.16$\pm0.03$.


\citet{poggianti06} proposed a scenario in which the population of
passive galaxies (those devoid of ongoing star formation at the time
they are observed) consists of two different components: primordial
passive galaxies (the most massive, mostly ellipticals), 
whose stars all formed at $z > 2-3$, and quenched
galaxies (on average less massive, mostly S0s), 
whose star formation has been truncated due to the dense
environment at later times.  Comparing with simulations, they found
that at $z=0$ the observed fraction of passive galaxies in clusters
resembles the fraction (in mass and in number of galaxies) that has
resided in clusters ($M_{sys} > 10^{14} M_{\odot}$) during at least the
last 3 Gyr. 

In this picture, the median fraction of
star-forming galaxies, and the median blue fraction we observe, do not
correlate with the cluster $\sigma$ or $L_X$. This is because in
clusters more massive than 500$\rm km \, s^{-1}$ at $z=0$, the median
fraction of galaxies that have spent enough time in a massive
environment to have their star formation quenched by some
(unidentified) environmental mechanism does not vary systematically
with cluster mass. This may be a viable explanation also for
the lack of correlation between the luminous-to-faint ratio and 
the global cluster properties, as the blue galaxies and the faint
red galaxies together form the overall faint cluster population
subject to quenching.

\subsection{Morphological fractions and red spirals}

When studying the CMD and, more generally, the evolution of galaxy
clusters, it is important to evaluate the fractions of
the different morphological types residing on the red
sequence. Indeed, apart from the well known presence of red Es in galaxy
clusters, it is not completely clear if there is some correlation
between the frequency of red S0s and late-type galaxies and the global
properties of the clusters.
The presence of S0s and late-type galaxies on the RS is often ascribed
to the capacity of the hostile cluster environment to quench star
formation or even strip and/or induce a fading of the disk of galaxies of 
the later types. 

\begin{figure*}
\centering
\includegraphics[scale=0.7]{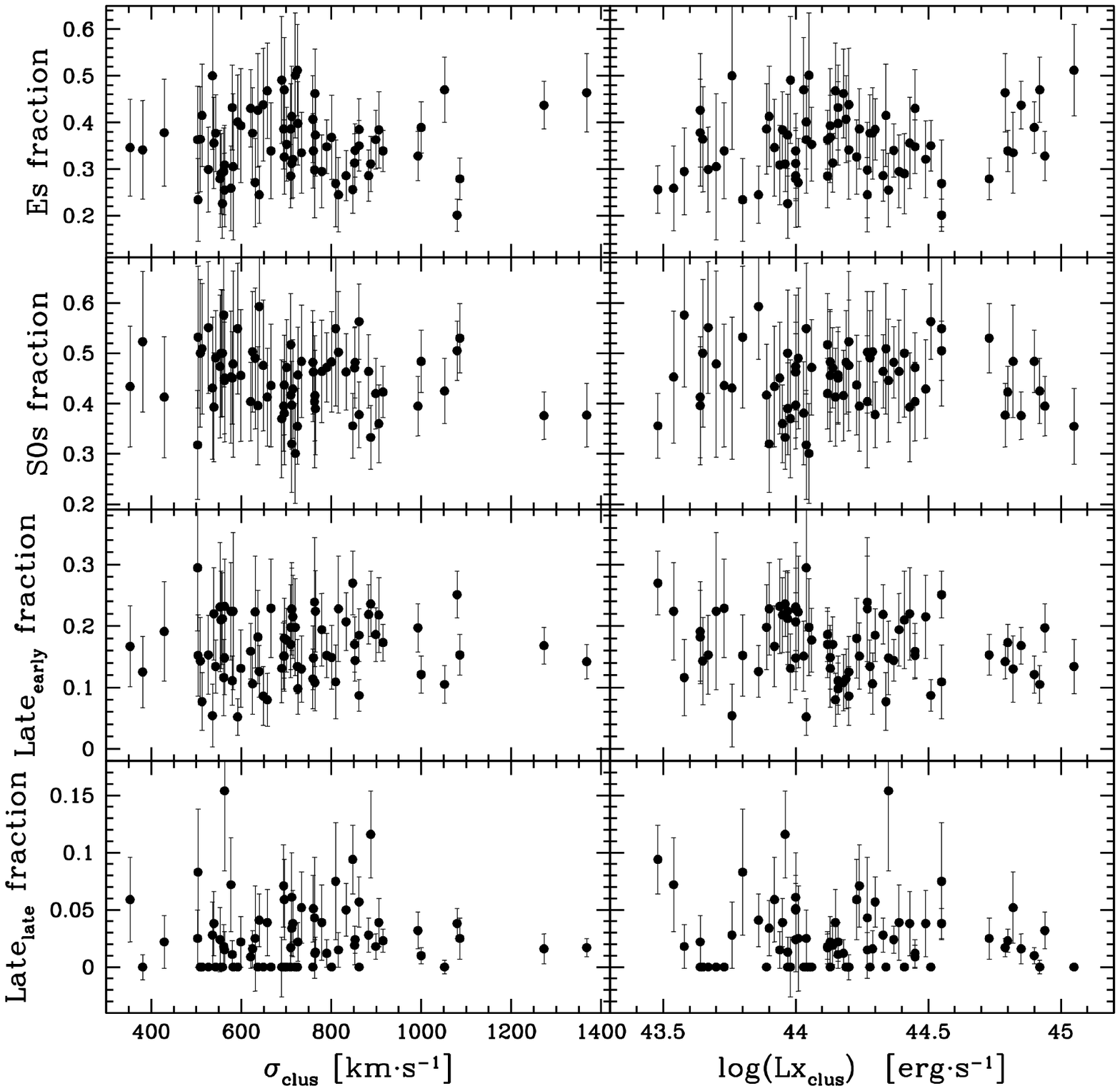}
\caption{Morphological fractions on the RS vs. central velocity dispersion and total X-ray luminosity of the WINGS clusters. Errorbars 
are the mean error on the 100 Montecarlo realizations. ``Early'' late-type galaxies are Sa's and Sb's, while ``Late'' late-type
galaxies are Sc's and later.
\label{fig:mortot}}
\end{figure*}

It is apparent from Fig.\ref{fig:mortot} that there is
no dependence of the RS morphological fractions on either X-ray
luminosity and cluster velocity dispersion. Similarly, no correlations
were found with the number of substructures, the BCG prevalence value
and the concentration of ellipticals (plot not shown). 
In the scenario of a
double channel for passive galaxies discussed above, this is expected
given that the median fractions of both primordial (ellipticals) and
quenched galaxies (S0s) are flat with the cluster velocity dispersion
(see Fig.~16 in \citet{poggianti06}).


The median morphological fractions of galaxies on the RS are
0.35$\pm$0.05 for ellipticals, 0.46$\pm$0.04 for S0s, 0.17$\pm$0.04
for early spirals (Sa to Sbc, $0<Ttype\leq 4$) and 0.02$\pm$0.02 for
later type spirals ($4 < T \leq 8$).  

It is interesting to compare these values with those found by
\citet{sb09} in clusters at $z=0.4-0.8$, which were obtained
adopting similar cluster radial limits and galaxy magnitude 
(passively evolved) limits to our study. These authors found 
an {\it increase} in the late-type fraction on the red sequence,
from 25\% at $z=0.75$ to 44\% at $z=0.45$, due to the fact that
the red-sequence becomes more populated at later times with disc
galaxies whose star formation has been quenched.
Interestingly, however, the same authors predict a subsequent
{\it decrease} of the red late-type fraction from $z=0.45$ to $z=0$,
due to the combination of the strong morphological evolution
from spirals to S0s and the mild variation in the RS luminous-to-faint
ratio observed in this redshift interval.

Indeed, our $\sim 20$\% fraction of late-type galaxies on the RS is much
lower than the value observed at $z=0.45$, and is consistent with 
a scenario in which over the interval $z=1-0.4$ 
the RS accretes spirals that have
stopped forming stars, a quite large fraction of which are transformed
into S0s at $z<0.4$ (Desai et al. 2007,
Dressler et al. 1997, Fasano et al. 2000, Postman
et al. 2005). 






\subsection{Elliptical to S0 ratios and cluster evolution}

Regarding the formation history of galaxy clusters, one may expect
that clusters with a large presence of red spirals may also have a
higher fraction of S0s compared to Es on the RS. In other words,
clusters with a larger number of passive late-type galaxies could have
transformed a larger number of them into S0s.

\begin{figure*}
\centering
\includegraphics[scale=0.7]{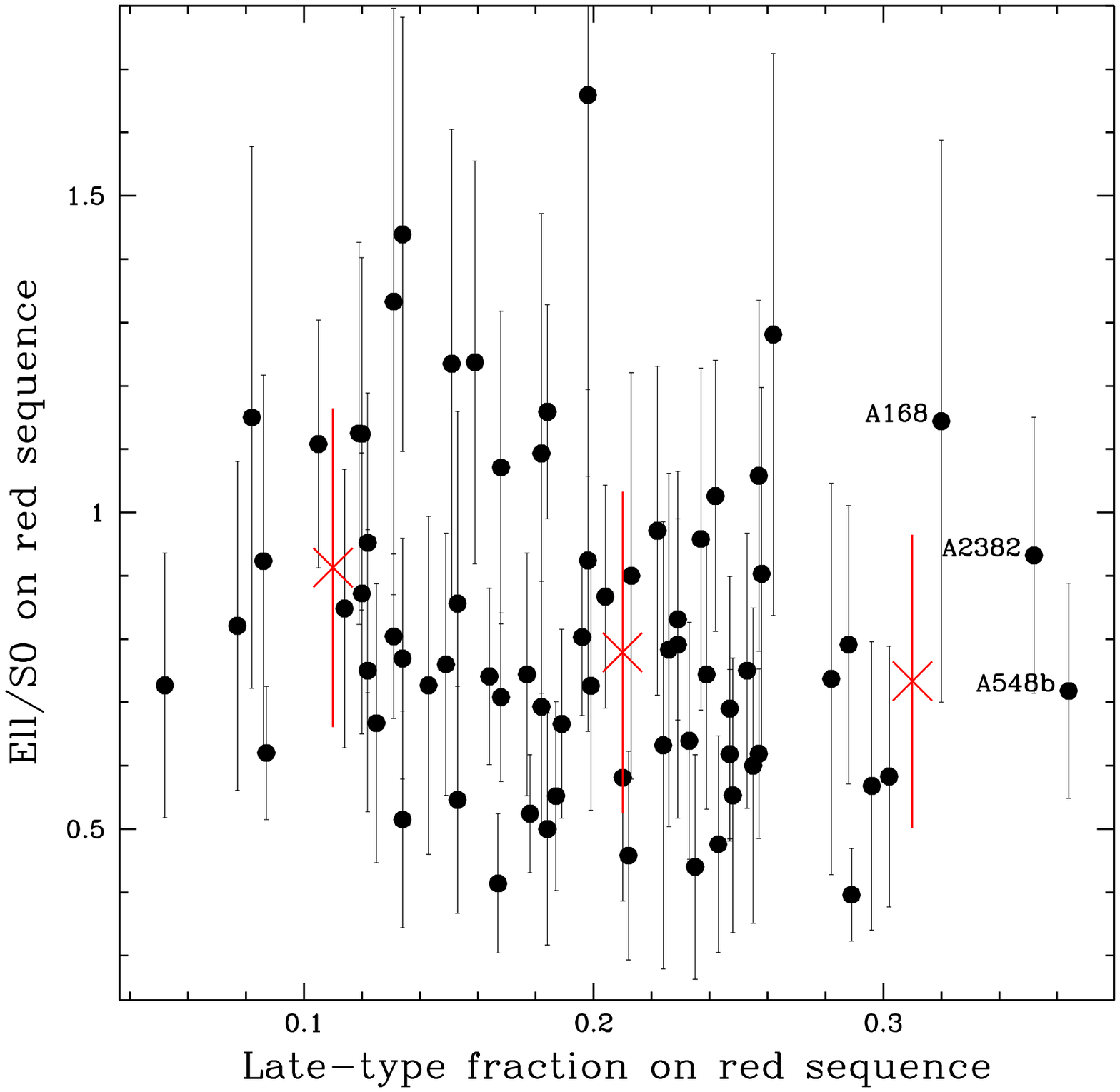}
\caption{Ellipticals over S0s number ratio on the RS 
vs. late-type fraction on the
RS. Errorbars are the mean error on the 100 Montecarlo
realizations. 
Red crosses and corresponding
errorbars represent the mean E/S0 values and their sigma in
three bins of late-type fraction.
\label{fig:es0}}
\end{figure*}

Figure\ref{fig:es0} shows the relation between the RS late-type fraction
and the RS ratio of Es to S0s. Clusters with a larger
fraction of red spirals tend to have a population of early-type galaxies on the
RS dominated by S0
morphologies. According to the Spearman test the E/S0 number ratio on
the RS is anticorrelated with the late type fraction on the RS with a
95\% probability.


\section{The main driver: the local density}

In the previous sections we have shown that the RS parameters, 
luminous-to-faint ratio, blue fraction and morphological fractions on the RS
do not depend on cluster global properties such as velocity dispersion
or X-ray luminosity, nor on elliptical galaxy concentration, BCG prevalence
 or number of substructures.

\begin{figure*}
\centering
\includegraphics[scale=0.7]{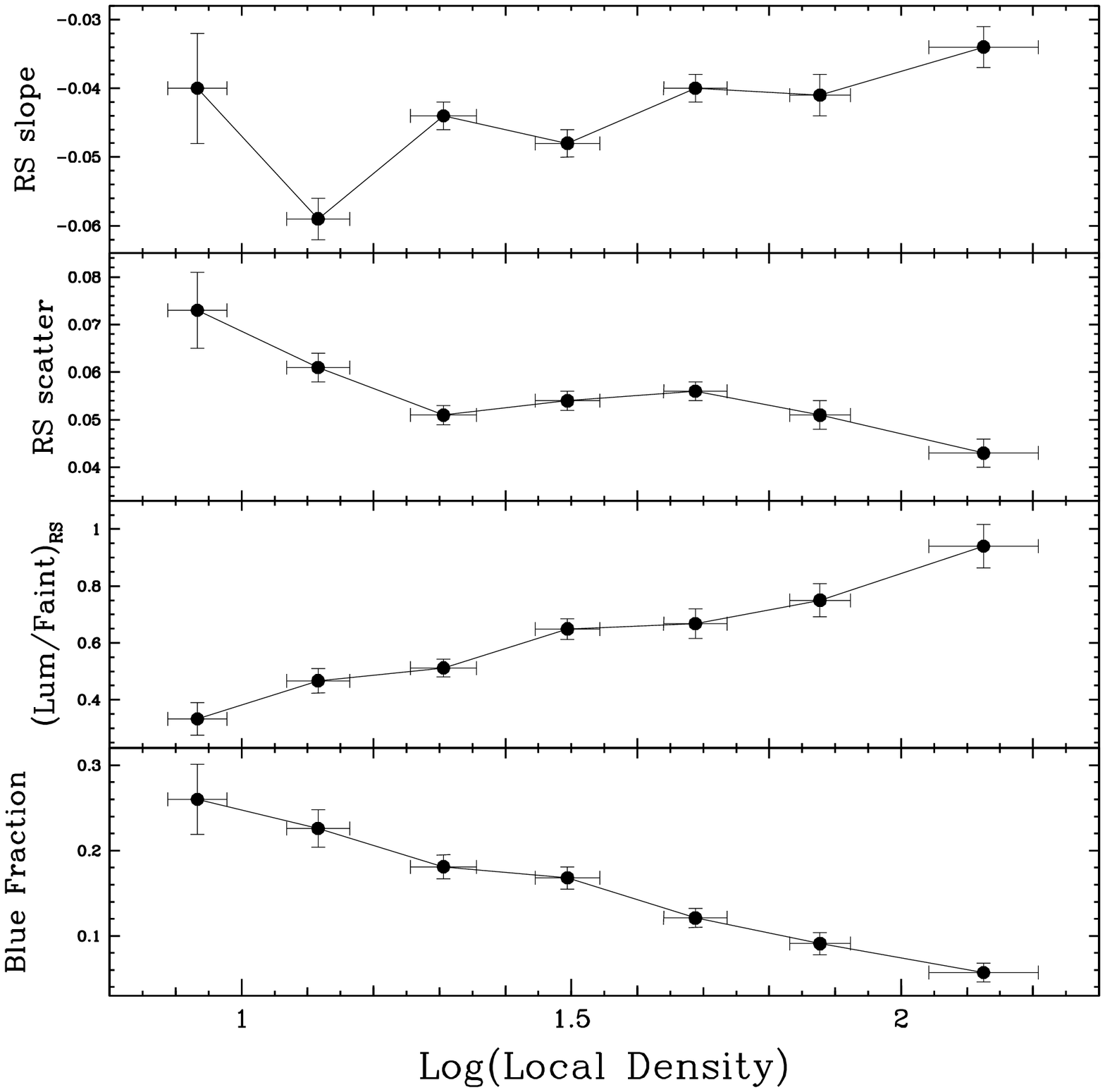}
\caption{RS slope and scatter, luminous-to-faint ratio and blue fraction
vs. galaxy local density.
\label{fig:ld1}}
\end{figure*}

\begin{figure*}
\centering
\includegraphics[scale=0.7]{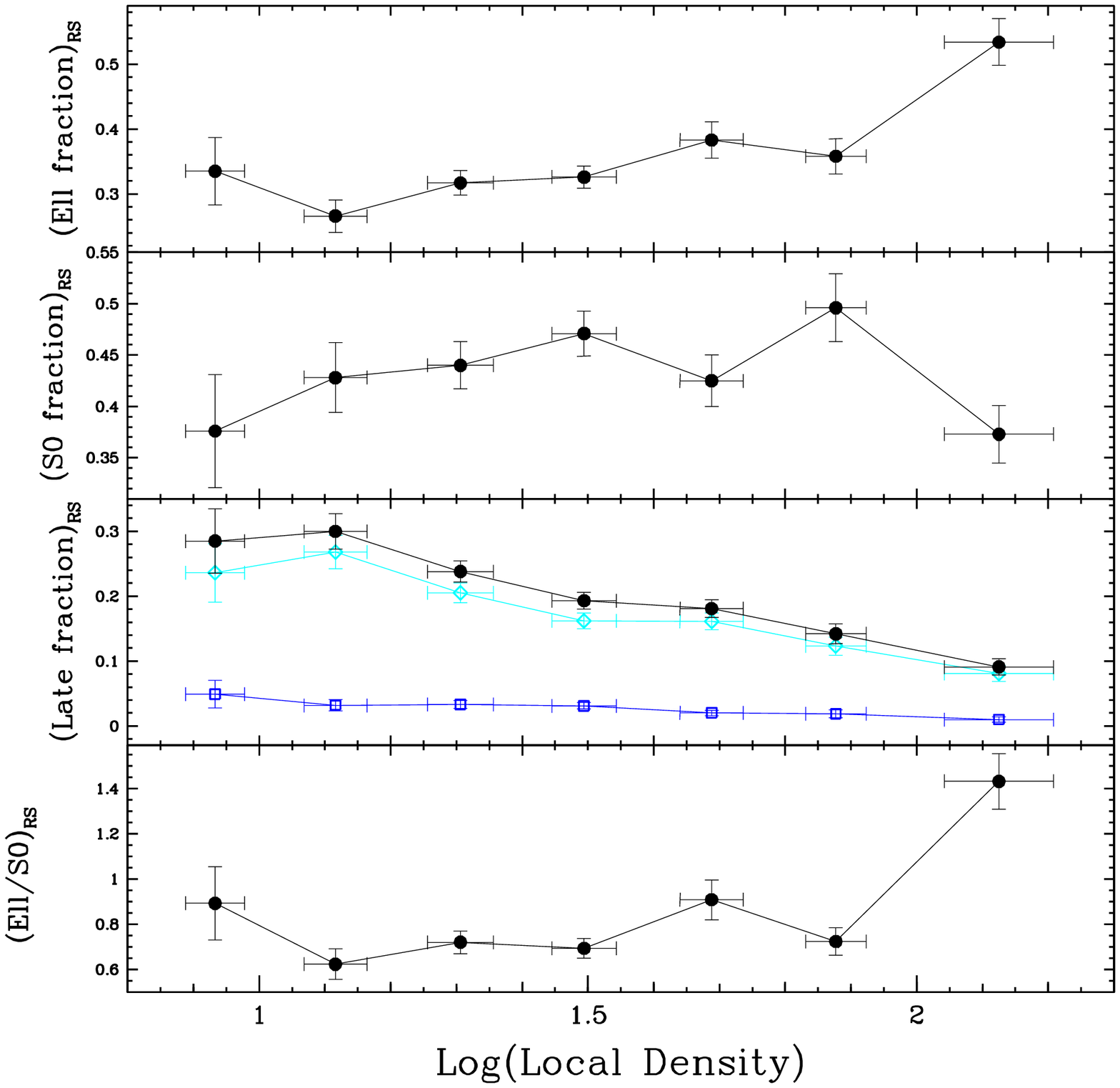}
\caption{Morphological fractions on the RS vs galaxy local density. From top
to bottom: ellipticals, S0s, later-types (early-spirals and late spirals, cyan
diamonds and blue squares, respectively) and E/S0 ratio.
\label{fig:ld2}}
\end{figure*}

In Figs.\ref{fig:ld1} and \ref{fig:ld2} we show that all the galaxy
properties considered do, however,
vary with the local galaxy density. Most noticeably,
higher density regions have a tighter RS (lower RS scatter),
and a lower fraction of blue galaxies. The RS generally becomes steeper (RS 
slope increases) with local density, except for a high value in the
lowest density bin. We note that a dependence of the location of
the CM relation and of the blue fraction
on local density has been found before in the Shapley supercluster
at z=0.05 by Haines et al. (2006).

The high RS slope in the lowest density bin is accompanied
by slighly higher values of the fraction of ellipticals
and of the E/S0 number ratio compared to denser regions (Fig.\ref{fig:ld2}). 
It is hard to assess the significance of these single points, and
wider images will be needed to sample lower density outer regions
of clusters.

Interestingly, the RS luminous-to-faint ratio increases monotonically
with local density (Fig.~8), 
implying that the relative proportion of luminous
red galaxies and faint red galaxies is higher in denser regions. This
is the first time such a trend is observed. At high-z, the opposite
trend has been found, consistent with the fact that the build-up of
the CMR is delayed in lower density environments \citep{tanaka05,
tanaka07}.  The increase of the luminous-to-faint ratio with density
in WINGS clusters seems to reflect, instead, the fact that WINGS
higher density regions host proportionally more high-mass than
low-mass galaxies (regardless of color) than lower density regions, as
shown in Vulcani et al. (2011).

%


Finally, higher density regions have a lower fraction of spirals
on the RS than lower density regions (Fig.\ref{fig:ld2}): the RS in the
highest density regions includes only a few spirals ($\sim 10$\%), while
the RS in low density regions is composed by up to 30\% of spirals.
The relative proportion of ellipticals and S0s 
does not change with local
density, except for the highest density regions where ellipticals dominate.

It is important to stress that the correlations between RS properties
and local density shown in Fig.\ref{fig:ld1} are {\it not} driven by
differences between the RS properties of the different morphological
types in combination with the morphology-density relation.
In particular, we wished to test whether a larger fraction of spirals
at low densities might induce the trends observed.
We found that none of the conclusions from Fig.\ref{fig:ld1}
change if we only include ellipticals, only S0s, or only 
ellipticals+S0s.

\section{Discussion and conclusions\label{sec:summary}}

In this work we have analyzed the properties of the (B-V) versus V
color-magnitude red sequence
of 72 X-ray selected clusters at $z=0.04-0.07$ from the WINGS survey.

We have searched for correlations between the characteristics of the
RS and the properties of both the global and local environment, finding
the following main results:

\begin{itemize}

\item The location and parameters of the RS, as well as the magnitude
and morphological distributions of its galaxies, do not depend on global 
cluster properties. Specifically:

\begin{itemize}
\item neither the slope nor the scatter of the RS, nor

\item the number ratio of luminous-to-faint galaxies on the RS, nor

\item the fraction of blue galaxies, nor

\item the morphological mix on the RS, that is, the fraction of galaxies
on the RS that are ellipticals, S0s and spirals,

\end{itemize}

none of the above vary systematically
with cluster velocity dispersion nor X-ray luminosity (hence with 
cluster mass), nor with the number of
substructures, BCG magnitude separation from the next brightest galaxies
nor spatial concentration of elliptical galaxies.

\item We find a weak tendency for clusters whose RS is ``rich'' in S0 galaxies
(compared to the population of ellipticals) to be also ``rich'' in red
spirals. The populations of S0 and spirals on the RS are coupled at some level.

\item 
The strongest correlations we find are between the RS and the local 
galaxy density. The scatter and, possibly, the slope of the RS
vary with local density, as do the luminous-to-faint ratio, blue galaxy 
fraction and morphological mix of galaxies on the RS.
The latter means that, remarkably, 
a clear morphology-density relation, especially for
spirals, is visible even restricting the analysis to galaxies on the 
red sequence.

\smallskip

However, the correlations between RS properties and local density are not
due to the morphology-density relation (i.e. a higher fraction of spirals
at lower densities), as all the trends in Fig.\ref{fig:ld1} 
persist even when restricting the analysis to only ellipticals, only S0s, 
or only early-type galaxies.

\end{itemize}

These trends suggest that galaxies in higher density regions 
within clusters became passive and evolved morphologically
towards earlier types at earlier epochs than galaxies in lower density
regions that are today {\it within the same clusters}, as witnessed by
the lower RS scatter, lower blue fraction,
and lower spiral fraction on the RS at higher densities.

Local density thus appears to be the main factor governing the
evolutionary pace of galaxies in these clusters. If this is the case,
the lack of a trend of the RS characteristics with the cluster mass is
not surprising, given that the distribution of local galaxy densities
does not depend on cluster mass \citep{poggianti10}.

In the same way as the morphological mix of galaxies (of all colors)
depends on local density \citep{dressler80} and not on cluster mass
\citep{poggianti09} in the local Universe, this work highlights the prominent
effect of the local density as opposed to the mass of the
galaxy host structure in setting the passivization epoch of galaxies and their
arrival on the RS, and the subsequent morphological transformation to S0s.

The dependence of the RS properties on local density can originate
from several mechanisms, and at different cosmological epochs. Since
the present local density is expected to correlate with the initial
density in the primordial phases, the star formation process may be
accelerated in dense regions from the start, yielding shorter
formation times.  Physical processes acting at later times, among
those commonly refereed to as environmental mechanisms, can accelerate
and/or quench star formation more efficiently in denser regions.
Among these, there are gravitational interactions between galaxies,
before or after the accretion of galaxies onto clusters, and gas
removal mechanisms.  While a detailed modeling of such processes is
beyond the scope of this paper, the combination of the lack of trends
with cluster mass and of the observed correlations with local density
represent a solid observational contraint for future studies.


\bibliography{biblioCMD}

\begin{acknowledgements}
  We thank the anonymous referee for her/his comments that led to improvements
  of this paper, and Diego Capozzi for useful discussions and comments.

  T.   Valentinuzzi   acknowledges a   post-doc  fellowship  from  the
  Ministero  dell'Istruzione,     dell'Universit\`a  e  della  Ricerca
  (Italy). BV and BMP acknowledge financial support from ASI contract 
  I/016/07/0. 
 	
  This  research has made use of  the NASA/IPAC Extragalactic Database
  (NED) which is operated by the Jet Propulsion Laboratory, California
  Institute of    Technology,   under  contract   with   the  National
  Aeronautics and Space Administration.

  IRAF (Image   Reduction and  Analysis   Facility)  is   written  and
  supported   by the IRAF programming  group   at the National Optical
  Astronomy Observatories (NOAO) in Tucson, Arizona.  NOAO is operated
  by the Association of Universities for Research in Astronomy (AURA),
  Inc.  under   cooperative   agreement with   the   National  Science
  Foundation.

\end{acknowledgements}

\end{document}